\documentclass[aps,pre,twocolumn,groupedaddress]{revtex4}
\usepackage{graphicx}
\usepackage{amsmath}
\begin{document}

\title{Accuracy thresholds of topological color codes on the hexagonal and square-octagonal lattices}
\author{Masayuki Ohzeki}
\homepage{http://www.stat.phys.titech.ac.jp/~mohzeki/english/}
\email{mohzeki@stat.phys.titech.ac.jp}
\affiliation{Department of Physics, Tokyo Institute of Technology, Oh-okayama, Meguro-ku,
Tokyo 152-8551, Japan}
\date{\today}

\begin{abstract}
Accuracy thresholds of quantum error correcting codes, which exploit topological properties of systems, defined on two different arrangements of qubits are predicted. 
We study the topological color codes on the hexagonal lattice and on the square-octagonal lattice by the use of mapping into the spin glass systems. 
The analysis for the corresponding spin glass systems consists of the duality, and the gauge symmetry, which has succeeded in deriving locations of special points, which are deeply related with the accuracy thresholds of topological error correcting codes. 
We predict that the accuracy thresholds for the topological color codes would be $1-p_c = 0.1096-8 $ for the hexagonal lattice and $1-p_c = 0.1092-3$ for the square-octagonal lattice, where $1-p$ denotes the error probability on each qubit. 
Hence both of them are expected to be slightly lower than the probability $1-p_c = 0.110028$ for the quantum Gilbert-Varshamov bound with a zero encoding rate.
\end{abstract}

\maketitle

\section{Introduction}
In quantum information processing, we encode information into a quantum state. 
However the quantum state suffers from decoherence by surroundings. 
We need an encoding technique to restore errors on the quantum state by decoherence, namely the error correcting code. 
It has been pointed out that error correction of the quantum state has a deep connection with spin glass models and lattice gauge theories in finite dimensions. 
An outstanding instance has been found between the random bond Ising model, as well as the random plaquette gauge model, and the topological toric code \cite{DKLP}.
Recently it has been shown that a more useful quantum error correcting code, the topological color code, also has a connection with the spin glasses \cite{BM,CC}.

\begin{figure}[tbp]
\begin{center}
\includegraphics[width=60mm]{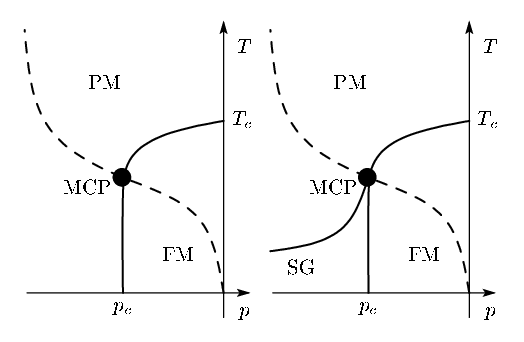}
\end{center}
\caption{{\protect\small Schematic picture of the phase diagram of the $\pm J$ Ising model on two-dimensional lattice (left panel) and on higher dimensions (right panel). 
The vertical axis expresses the temperature $T$, and the horizontal line denotes the concentration $p$ of the antiferromagnetic interactions. 
The multicritical point (MCP) is described by the black point. 
The Nishimori line is drawn by the dashed line. For higher dimensions, not only the ferromagnetic (FM) and paramagnetic phases (PM) but also the spin glass phase (SG) exists. 
The phase boundary under the Nishimori line is believed to be vertical by the argument of the gauge transformation \protect\cite{Rev3}.}}
\label{fig1}
\end{figure}

These fascinating bridges between the spin glasses and the quantum error correcting codes are established in a special subspace in spin glasses. 
This subspace is known as the Nishimori line which goes through the phase boundary between the ferromagnetic and paramagnetic phases as shown in Fig. \ref{fig1} \cite{Rev3,HN81}. 
The ferromagnetic phase on the Nishimori line represents the region where errors on the quantum state can be restored, and vice versa. 
Thus the critical point on this subspace in spin glasses corresponds to an accuracy threshold of the quantum error correcting code.
The critical point on this line is called the multicritical point. 
There are little exact and rigorous approaches for spin glasses in finite dimensions. 
One of the powerful theoretical techniques is the gauge symmetry on the Nishimori line. 
On this line, we can calculate the exact value of the internal energy, evaluate the upper bound for the specific heat, and obtain some sets of rigorous and exact inequalities by the gauge symmetry \cite{Rev3,HN81}. 
In addition, the analytical theory estimating the precise location of the multicritical point has been established by a technique with the duality and the replica method \cite{NN,MNN,TN,ONB,Ohzeki}. 
In the present study, we analyze two spin glass models, the random three-body Ising model ($\pm J$ type) on the triangular lattice and the Union-Jack lattice.
These models have deep connections with the topological color codes defined on the hexagonal and square-octagonal lattices, which implements several useful quantum gates \cite{BM,CC}. 
We present the analysis of estimations of the error thresholds of these topological color codes through the analysis of the location of the multicritical point for the random three-body Ising model on two lattices. 
As a result, we predict the accuracy threshold of the color code on the hexagonal lattice, which corresponds to the random three-body Ising model on the triangular lattice, would be $1-p_{c}=0.1096-8$ and that on the square-octagonal lattice, which has a connection with the random three-body Ising model on the Union-Jack lattice, would be $1-p_c =0.1092-3$, where $1-p$ represents the error probability on qubits by decoherence. 
\begin{figure}[tb]
\begin{center}
\includegraphics[width=50mm]{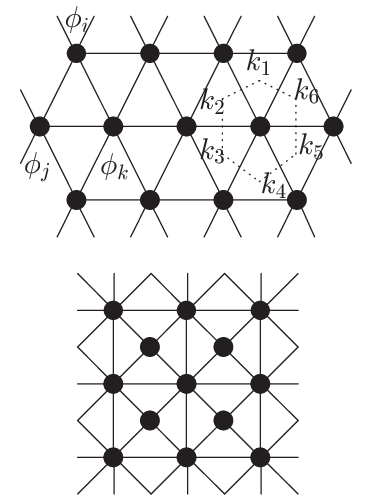}
\end{center}
\caption{{\protect\small Triangular lattice and Union-Jack lattice. The site variable $\protect\phi$ is located at each site, and the plaquette variable $k_{I}$ is introduced at each triangular face. 
The plaquette variables form a structure of the hexagonal and square-octagonal lattices.}}
\label{fig2}
\end{figure}

\section{Duality}
Let us review several known facts in the present section to fix the notation and prepare for the developments in the next section. 
We take the three-body Ising model on the triangular lattice as an example here for simplicity. 
It is straightforward to generalize the following arguments to the Union-Jack lattice.

The Hamiltonian of the three-body Ising model on the triangular lattice is 
\begin{equation}
H=-J\sum_{\langle ijk\rangle }\sigma _{i}\sigma _{j}\sigma _{k},
\end{equation}%
where the summation is taken over all the triangles (both of up-pointing and down-pointing triangles), and $\sigma $ is the Ising spin variable taking $\pm 1$ on sites as in Fig. \ref{fig2}. 
This model is known as a self-dual model \cite{3SD1,3SD2}, and has a single transition point $T_{c}=2.26919$ which is the same as the ordinary two-body Ising model on the square lattice \cite{BW}. 
We use another expression for convenience as follows, 
\begin{equation}
H=-J\sum_{\langle ijk\rangle }\cos \left\{ \pi (\phi _{i}+\phi _{j}+\phi_{k})\right\}.
\end{equation}%
Here the spin variable is changed into $\phi $ taking $0$ and $1$. 
It will be useful to review the duality transformation for the non-random three-body Ising model on the triangular lattice through the Wu-Wang formalism \cite{WuWang}, which is more suitable for application to spin glasses. 
We consider the face Boltzmann factor for an elementary triangle, 
\begin{equation}
A_{\phi _{i}+\phi _{j}+\phi _{k}}=\mathrm{e}^{K\cos \left\{ \pi
(\phi_{i}+\phi _{j}+\phi _{k})\right\} },
\end{equation}%
where $K$ is the coupling constant $K=\beta J$. We introduce two-component Fourier transform of the face Boltzmann factor as, 
\begin{equation}
A_{k }^{\ast }=\frac{1}{\sqrt{2}}\left( \mathrm{e}^{K}+\mathrm{e}^{-K}\cos\pi k \right).
\end{equation}%
If we use this quantity and another variable $k_{I}$, which takes $0$ or $1$, on each elementary triangle as in Fig. \ref{fig2}, we can rewrite the face Boltzmann factor as, similarly to case for the ordinary two-body Ising model on the square lattice \cite{WuWang}, 
\begin{equation}
A_{\phi _{i}+\phi _{j}+\phi _{k}}=\frac{1}{\sqrt{2}}\left[\sum_{k_{I}=0,1}A_{k_{I}}^{\ast }\mathrm{e}^{\mathrm{i} \pi k_{I}(\phi_{i}+\phi _{j}+\phi _{k})}\right].
\end{equation}%
Using this expression, we can rewrite the partition function of the three-body Ising model on the triangular lattice as, 
\begin{eqnarray}
Z(K) &=& \sum_{\{\phi_{i}\}} \prod_{I} A_{\phi _{i}+\phi _{j}+\phi _{k}}
\label{d1} \\
& = & \left( \frac{1}{\sqrt{2}}\right) ^{N}\sum_{\{\phi
_{i}\}}\sum_{\{k_{I}\}}\prod_{I}A_{k_{I}}^{\ast }\mathrm{e}^{\mathrm{i} \pi
k_{I}(\phi _{i}+\phi _{j}+\phi _{k})},
\end{eqnarray}
where $N$ is the number of triangles. 
We take the summation over $\phi $, and obtain constraints expressed by the Kronecker delta with modulus $2$ on each site $i$ on the triangular lattice $2\delta(k_{1}+k_{2}+k_{3}+k_{4}+k_{5}+k_{6})$, where $k_{I}$ surrounds each site as in Fig. \ref{fig2}. 
The partition function is then rewritten as, 
\begin{eqnarray}
Z(K) &=&2^{N/2}\left( \frac{1}{\sqrt{2}}\right)^{N}\sum_{\{k_{I}\}}\prod_{I}A_{k_{I}}^{\ast }  \notag \\
&&\quad \times \prod_{i}\delta (k_{1}+k_{2}+k_{3}+k_{4}+k_{5}+k_{6}).
\end{eqnarray}
We introduce another variable $k_{I}=\varphi _{i}+\varphi _{j}+\varphi _{k}$(mod $2$), which satisfies all the constraints, and obtain 
\begin{equation}
Z(K)=\sum_{\{\varphi _{i}\}}\prod_{i}A_{\varphi _{1}+\varphi _{2}+\varphi_{3}}^{\ast },  \label{d2}
\end{equation}%
where $\varphi _{i}$ also takes $0$ and $1$ and is located at each site. 
We obtain two expressions for the partition function $Z(K)$ by the original and dual face Boltzmann factors as in Eqs. (\ref{d1}) and (\ref{d2}). 
We extract the principal Boltzmann factors $A_{0}$ and $A_{0}^{\ast }$ as, to measure the energy for both of these expressions from the state with all-spin up, 
\begin{equation}
A_{0}^{N}z(u_{1})=A_{0}^{\ast N}z(u_{1}^{\ast }),
\end{equation}%
where $z$ stands for the normalized partition function $z(u_1)=Z/A^N_0$ and $z(u^*_1)=Z/(A_0^{*})^N$. 
The quantities $u_{1}$ and $u_{1}^{\ast }$ are defined as $u_{1}=A_{1}/A_{0}$ and $u_{1}^{\ast}=A_{1}^{\ast }/A_{0}^{\ast }$. 
Each partition function is now reduced to a single-variable function of $u_{1}=\mathrm{e}^{-2K}$ and $u_{1}^{\ast}=\tanh K$. 
We can identify the critical point as a fixed point of the duality by solving $\mathrm{e}^{-2K_{c}}=\tanh K_{c}$, under the assumption of a unique transition. 
We remark that, at this fixed point, an appealing equation for two principal Boltzmann factors is satisfied, $A_{0}=A_{0}^{\ast }$.

\section{Multicritical Point}
We step in the analysis of the location of the multicritical point for the random three-body Ising model on the triangular lattice, which is related with the accuracy threshold of the topological color code on the hexagonal lattice \cite{CC}. 
The Hamiltonian is slightly modified as 
\begin{equation}
H=-J\sum_{\langle ijk\rangle }\tau _{ijk}\cos(\phi_{i}+\phi_{j}+\phi_{k}),
\end{equation}
where $\tau _{ijk}$ denotes the quenched random coupling. 
Though various types of distribution for $\tau _{ijk}$ can be considered, we here restrict ourselves to the $\pm J$ Ising model, which has a connection with the topological color code. 
The distribution function for the $\pm J$ Ising model is 
\begin{equation}
P(\tau _{ijk})=p\delta (\tau _{ijk}-1)+(1-p)\delta (\tau _{ijk}+1)=\frac{%
\mathrm{e}^{K_{p}\tau _{ijk}}}{2\cosh K_{p}},
\end{equation}
where $K_{p}$ is defined by $\mathrm{e}^{-2K_{p}}=(1-p)/p$. 
The Nishimori line is given by the condition $K=K_{p}$ \cite{HN81,Rev3}.

We apply the replica method to the $\pm J$ Ising model. 
The $n$-replicated partition function after the configurational average is 
\begin{equation}
Z_{n}=\left[ \sum_{\{\sigma _{i}\}}\prod_{\langle ijk\rangle }\prod_{\alpha
=1}^{n}\exp \left\{K\tau _{ijk}\sigma _{i}^{\alpha
}\cos(\phi_{i}+\phi_{j}+\phi_{k})\right\}\right] _{\mathrm{av}},
\end{equation}
where $n$ stands for the replica number and the angular brackets denote the configurational average. 
We generalize the duality argument to the $n$-replicated $\pm J$ Ising model \cite{NN,MNN,TN,ONB,Ohzeki}. 
For this purpose it is useful to define the face Boltzmann factor $A_k ~(k =0,1,\cdots, n)$, which represents the configuration-averaged face Boltzmann factor for interacting spins with $k$ triplets giving $\phi_i + \phi_j+\phi_k = 1$ (mod $2$) among $n$ triplets for a triangle. 
The duality gives the relationship of the partition functions with different values of the face Boltzmann factor as given by 
\begin{eqnarray}
&&Z_{n}(A_{0},A_{1},\cdots ,A_{n})  \notag \\
&&\quad =Z_{n}(A_{0}^{\ast },A_{1}^{\ast },\cdots ,A_{n}^{\ast }),
\label{PF}
\end{eqnarray}
The dual face Boltzmann factors $A_k^*$ are defined by the discrete multiple Fourier transforms of the original face Boltzmann factors, which are simple combinations of plus and minus of the original Boltzmann factors as the non-random case shown above. 
Two principal Boltzmann factors are important pieces of information and given as $A_{0}=2\cosh\left\{ \left(n+1\right)K\right\} $, and $A_{0}^{\ast }=2^{n/2}\cosh ^{n}K$, similarly to the case for the random bond Ising model on the square lattice \cite{NN,MNN}. 
We extract these principal face Boltzmann factors from Eq. (\ref{PF}) to measure the energy from the all-parallel spin configuration, 
\begin{eqnarray}
&&{A_{0}}^{N}z_{n}(u_{1},u_{2},\cdots ,u_{n})  \notag \\
&&\quad ={A_{0}^{\ast }}^{N}z_{n}(u_{1}^{\ast },u_{2}^{\ast },\cdots
,u_{n}^{\ast }).  \label{PF1}
\end{eqnarray}
where $z_n(u_1,\cdots)$ and $z_n(u^*_1,\cdots)$ are defined as $Z_n/A^{N}_0$ and $Z_n/(A^{*}_0)^{N}$.

The duality identifies the critical point under the assumption of a unique phase transition. 
It is given as the fixed point of the duality and is known to yield the exact critical point for a simple ferromagnetic system as reviewed in the previous section. 
In order to obtain the multicritical point of the present replicated spin glass system, we set $K=K_p$, which defines the Nishimori line on which the multicritical point is expected to lie.
Since $z_n$ is a multivariable function, there is no fixed point of the duality relation in the strict sense which satisfies $n$ conditions simultaneously, $u_1(K)=u^*_1(K),u_2(K)=u^*_2(K),\cdots,u_n(K)=u^*_n(K)$.
This is in sharp contrast to the non-random case, in which the duality is a relation between single-variable functions. 
We nevertheless set a hypothesis that a single equation $A_{0}=A_{0}^{\ast }$ gives the location of the multicritical point for any replica number $n$ \cite{NN,MNN}, similarly to the non-random three-body Ising model at the critical point. 
The quenched limit $n\rightarrow 0$ for the equation $A_{0}=A_{0}^{\ast }$ yields \cite{NN,MNN}, 
\begin{equation}
-p\log p-(1-p)\log (1-p)=\frac{1}{2}\log 2.
\end{equation}%
The solution to this equation is $p_{c}=0.889972$. 
This estimation $p_{c}=0.889972$ is equivalent to the probability for the quantum Gilbert-Varshamov bound with zero rate encoding $1-p_c = 0.110028$ \cite{GV1,GV2}, and consistent with the very recent numerical estimation $p_{c}=0.891(2)$ by Katzgraber {\it et al.} \cite{CC}.

For the random three-body Ising model on the Union-Jack lattice, the same duality analysis can be applied and we obtain the same solution. 
In the context of the quantum error correction, these results mean that the accuracy thresholds for the color codes on the hexagonal and square-octagonal lattices are given by $1-p_{c}=0.110028$ for error probability on qubits. 
This consequence would not be correct, since two color codes have different computational capabilities \cite{BM2}. 
For the topological color code on the square-octagonal lattice, it is possible to implement the whole Clifford group of unitary gates generated by the Hadamard gate, the $\{\pi/8\}$ gate, and the controlled-NOT gate; while, for that on the hexagonal lattice, the $\{\pi/8\}$ gate cannot be implemented.
In spin glasses, the quenched randomness yields frustration depending on geometry of lattices, and the locations of the multicritical points in quenched random systems are thus different in general. 
In addition, the symmetry (e.g., spin-reversal) is quite different between two-body and three-body Ising models, while the above estimation is also given for the case of the two-body $\pm J $ Ising model on the self-dual lattices \cite{NN,MNN}. 
The above naive approach by the analysis of two face Boltzmann factors defined on a single triangle would not yield correct answer which reflects on the detailed property of systems, the symmetry and the shape of the lattice under consideration. 
Indeed the detailed analysis through exact calculations by the virtue of the hierarchical lattice has shown deviations between the results obtained by the above naive approach and the answers \cite{ONB,HB}. 
It is, however, very difficult to obtain a precise value of the location of the multicritical point, that is the accuracy threshold, by numerical approaches because of the necessity of very long time for the equilibration and statistical error for the average of the quenched randomness.
In the following section, instead of the numerical approach, we apply an analytical theory, the improved method for a precise location of the multicritical point \cite{Ohzeki}. 
The purpose of the following analysis is to find deviations from the $p_{c}=0.889972$ given by the naive approach and to detect a slight difference between the location of the multicritical points for the triangular and Union-Jack lattices.

\section{Improved method}
\begin{figure}[tb]
\begin{center}
\includegraphics[width=50mm]{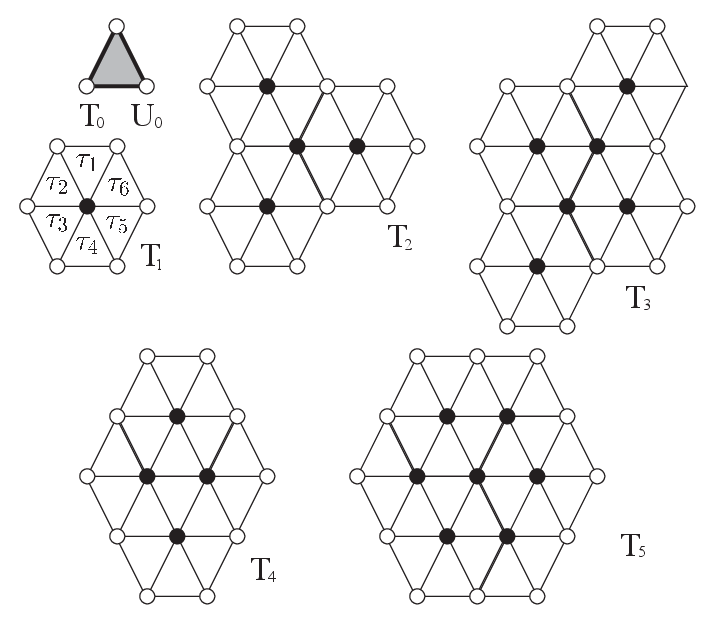}
\end{center}
\caption{{\protect\small Several clusters for the $\pm J$ three-body Ising
model on the triangular lattice. These are chosen to cover the whole
lattice. A cluster is the unit plaquette encircled by white spins. The spins
marked black on the original lattice are traced out to yield interactions
among white spins in clusters. The case denoted by $T_0$ and $U_0$ gives the principal
Boltzmann factor used in the naive approach.}}
\label{fig3}
\end{figure}

\begin{figure}[b]
\begin{center}
\includegraphics[width=45mm]{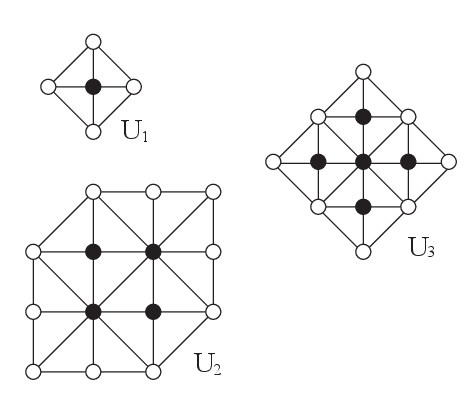}
\end{center}
\caption{{\protect\small Several clusters for the $\pm J$ three-body Ising
model on the Union-Jack lattice. }}
\label{fig4}
\end{figure}

As shown in Figs. \ref{fig3} and \ref{fig4}, let us consider to sum over a part of the spins, called a cluster below, on the triangular lattice and the Union-Jack lattice. 
The previous technique is only consideration of the \textit{"local"} principal Boltzmann factors $A_{0}$ and $A_{0}^{\ast }$ defined on an elementary triangle, which does not necessarily reflect the effects of frustration inherent in spin glasses. 
\begin{figure}[b]
\begin{center}
\includegraphics[width=50mm]{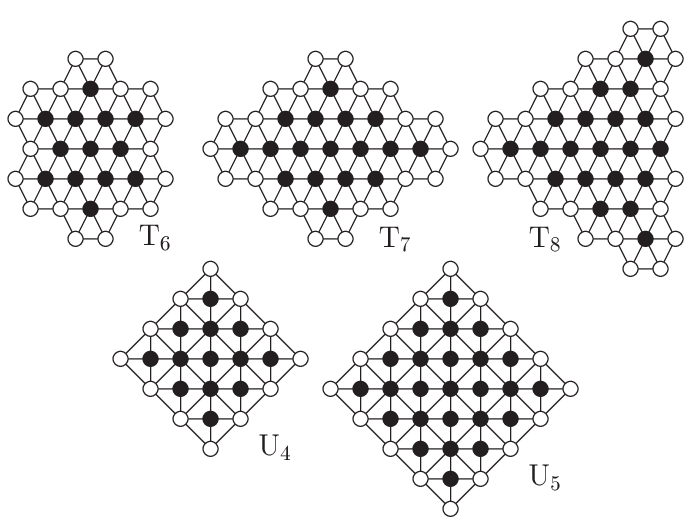}
\end{center}
\caption{{\protect\small Supplementary clusters for the computation of the slope of $T_c$ at $p=1$.}}
\label{fig5}
\end{figure}
Then the set of clusters must be chosen to cover the whole lattice under consideration as in Figs. \ref{fig3} and \ref{fig4}, where six examples for the triangular lattice and three ones for the Union-Jack lattice of the choices of clusters are depicted. 
It is expected that, if we deal with clusters of larger sizes, the improved method shows systematic improvements toward the exact answer on the location of the multicritical point \cite{Ohzeki}.

Then the duality relation for the replicated partition functions is reduced to, 
\begin{eqnarray}
&&Z_{n}^{(r)}(A_{0}^{(r)},A_{1}^{(r)},\cdots )  \notag \\
&&\quad =Z_{n}^{(r)}(A_{1}^{\ast (r)},A_{2}^{\ast (r)},\cdots ),  \label{PF2}
\end{eqnarray}%
where the superscript $r$ stands for the type of the cluster that one chooses. 
The quantity $A_{k}^{(r)}$ is the local Boltzmann factor including many-body interactions generated by summation over spins marked black in Fig. \ref{fig3}. 
We define the principal Boltzmann factors $A_{0}^{(r)}$ and its dual $A_{0}^{\ast (r)}$ as those with all spins surrounding the cluster in the up state. 
We assume that a single equation gives the accurate location of the multicritical point for any number of $n$, similarly to the naive approach, 
\begin{equation}
A_{0}^{(r)} = A_{0}^{\ast (r)}.  \label{Improved}
\end{equation}%
This is the improved method to predict a location of the multicritical point with higher precision than the naive approach.

One example is the hexagonal cluster as denoted by $\mathrm{T_1}$ in Fig. \ref{fig3}. 
The principal Boltzmann factors of the type $\mathrm{T_1}$ are given as 
\begin{eqnarray}
A_{0}^{(\mathrm{T_1})} &=&\left[ \sum_{\{\sigma _{0}^{\alpha
}\}}\prod_{\alpha =1}^{n}\prod_{I=1}^{6}\mathrm{e}^{K\tau _{I}\sigma
_{0}^{\alpha }}\right] _{\mathrm{av}}  \notag \\
&=&\left[ 2^{n}\cosh ^{n}\left( \sum_{I=1}^{6}K\tau _{I}\right) \right] _{%
\mathrm{av}}  \label{LPB1}
\end{eqnarray}
and 
\begin{eqnarray}
A_{0}^{\ast (\mathrm{T_1})} &=&\left[ 2^{-3n}\sum_{\{\sigma _{0}^{\alpha
}\}}\prod_{\alpha =1}^{n}\prod_{I=1}^{6}\left( \mathrm{e}^{K \tau_{I}}+%
\mathrm{e}^{-K \tau_{I}}\sigma _{0}^{\alpha }\right) \right] _{\mathrm{av}} 
\notag \\
&=&\left[ 2^{3n}\left( \cosh ^{6}K +\prod_{I=1}^{6}\tau_I \sinh^6 K \right) %
\right] _{\mathrm{av}},  \label{LPB2}
\end{eqnarray}%
where $\tau _{I}$ is defined on each plaquette as shown in Fig. \ref{fig3}.
We estimate a more precise location of the multicritical point of the $\pm J$ three-body Ising model on the triangular lattice as $p_{c}=0.890212$ by Eq. (\ref{Improved}).
\begin{table}[t]
\begin{center}
\begin{tabular}{lllll}
\hline
Type & \# & $p_c$ & $1 - p_c$ & Slope \\ \hline\hline
$\mathrm{T_0,U_0}$ & $1$ & $0.889972$ & $0.110028$ & $3.41421$ \\ \hline
$\mathrm{T_1}$ & $6$ & $0.890212$ & $0.109788$ & $3.39884$ \\ 
$\mathrm{T_2}$ & $18$ & $0.890321$ & $0.109679$ & $3.39031$ \\ 
$\mathrm{T_3}$ & $24$ & $0.890358$ & $0.109617$ & $3.38478$ \\ 
$\mathrm{T_4}$ & $16$ & $0.890344$ & $0.109656$ & $3.38507$ \\ 
$\mathrm{T_5}$ & $24$ & $0.890404$ & $0.109596$ & $3.37558$ \\ \hline
$\mathrm{T_6}$ & $42$ & none & none & $3.36985$ \\ 
$\mathrm{T_7}$ & $54$ & none & none & $3.36796$ \\ 
$\mathrm{T_8}$ & $60$ & none & none & $3.36736$ \\ \hline
$\mathrm{U_1}$ & $4$ & $0.890725$ & $0.109275$ & $3.33658$ \\ 
$\mathrm{U_2}$ & $16$ & $0.890394$ & $0.109606$ & $3.37078$ \\ 
$\mathrm{U_3}$ & $16$ & $0.890825$ & $0.109075$ & $3.32365$ \\ \hline
$\mathrm{U_4}$ & $36$ & none & none & $3.30885$ \\ 
$\mathrm{U_5}$ & $64$ & none & none & $3.29998$ \\ \hline\hline
Perturbation &  &  &  & $3.20911^{\rm a}$ \\ \hline
\end{tabular}
\end{center}
\begin{flushleft}
$^{\rm a}$References \cite{TRSlope} and \cite{Domany}.
\end{flushleft}
\caption{{\protect\small Location of the multicritical point of the $\pm J$ three-body Ising model on the triangular lattice (from $\mathrm{T_1}$ to $\mathrm{T_5}$) and on the Union-Jack lattice (from $\mathrm{U_1}$ to $\mathrm{U_3}$). 
The number of the triangles in the cluster is written in the second column. 
We estimate the values for the slope at the critical point $T_{c}$ and set them in the rightmost column.}}
\label{result}
\end{table}

One needs to calculate two local principal Boltzmann factors by summation over the internal spins, and evaluate the configurational-averaged values for them as in Eqs. (\ref{LPB1}) and (\ref{LPB2}) for the improved method. 
The computational complexity of the estimation of the multicritical point by the improved method is $O(2^{N^{(r)}+N_{s}^{(r)}})$, where $N^{(r)}$ is the number of interactions in the cluster and $N_{s}^{(r)}$ is that of internal sites. 
In reasonable time, we perform the improved method for the location of the multicritical point on the triangular lattice from $N^{(\mathrm{T_{1}})}=6$ to $N^{(\mathrm{T_{5}})}=24$, and on the Union-Jack lattice from $N^{(\mathrm{U_{1}})}=4$ to $N^{(\mathrm{U_{3}})}=16$ as in Figs. (\ref{fig3}) and (\ref{fig4}). 
The results obtained by the improved method are summarized in Table \ref{result}. 
As you can see, all the results by the improved method show higher values $p_{c}$ than $p_{c}=0.889972$ by the naive approach. 
In the context of the quantum error correcting code, the error threshold of the topological color code is lower than the probability $1-p_{c}=0.110028$ for the quantum Gilbert-Varshamov bound with zero rate encoding as in Table \ref{result} \cite{GV1,GV2}.

Next let us to examine the performance of the improved method, how close to the exact answer, for the random three-body Ising model on the triangular and Union-Jack lattices. 
If we formulate the above analysis without restriction to the Nishimori line, we can give an approximative shape of the phase boundary \cite{Ohzeki}. 
We estimate the value of the slope at the critical point $T_{c}$ for $p=1$ as listed in Table \ref{result}. 
It is one of the references for the performance of the improved method to compare these values with the solution obtained by a simple perturbation $3.20911$ \cite{TRSlope}, which is essentially the same way as Domany's result for the two-body Ising model on the square lattice \cite{Domany}. 
Since the computational complexity can be reduced to $O(N^{(r)}2^{N_{s}^{(r)}})$ for the calculation of the slope of $T_{c}$ at $p=1$ by the improved method, we evaluate only the values of the slope by using larger size of the clusters denoted by $\mathrm{T_{6}}$, $\mathrm{T_{7}}$, and $\mathrm{T_{8}}$ as in Fig. \ref{fig5}. 
As you can see in Table. \ref{result} for the case for the triangular lattice, if the number of the interactions in the cluster becomes larger, the estimations for the slope of $T_{c}$ by the improved method steadily decrease into the perturbation solution $3.20911$ but the convergence into the value is extremely slow.
Such a behavior of the slope of $T_{c}$ at $p=1$ is a sharp difference from the case for the two-body $\pm J$ Ising model on the square lattice \cite{Ohzeki}. 
Though it is thus expected that the predictions of the location of the multicritical point by the improved method certainly approach the exact answer, our estimation $p_{c}=0.8902-4$ for the random three-body Ising model on the triangular lattice is a delicate conclusion. 
At least, we can state that the accuracy threshold $1-p_{c}$ for the color code on the hexagonal lattice is lower than the quantum Gilbert-Varshamov bound with a zero rate encoding.

On the other hand, we obtain a relatively reliable result for the Union-Jack lattice, though we find an exception of the systematic improvement by the use of larger size clusters the result by ${\rm U_{2}}$ cluster. 
From the difference between two values for the slope of $T_{c}$ at $p=1$ by the improved method and the perturbation solution, ${\rm U_{2}}$ cluster is considered to be an wrong approximation for estimations of the location of the multicritical point. 
It is considered that the shape of $\mathrm{U_{2}}$ cluster is slightly different from the series of other clusters as shown in Fig. \ref{fig4}. 
If we use a series of the similar clusters to $\mathrm{U_{1}}$ and $\mathrm{U_{3}}$ (see Fig. \ref{fig5}) for the further approximations for the slope of $T_{c}$ at $p=1$, we find convergence into the perturbation result $3.20911$, which can be derived in the same way in Refs. \cite{TRSlope} and \cite{Domany}, with a similar degree to the case for the two-body $\pm J$ Ising model on the square lattice \cite{Ohzeki}. 
We thus predict that the location of the multicritical point for the Union-Jack lattice would be $p_{c}=0.8907-8$ from the results by $\mathrm{U_{1}}$ and $\mathrm{U_{3}}$ clusters, or equivalently the accuracy threshold for the color code on the square-octagonal lattice is $1-p_{c}=0.1092-3$.

The improved method thus strongly depends on the shape of the used cluster, as well as the number of interactions included in the cluster. 
It is considered that the systematic improvement with convergence into the answer is yielded by the ingenious choice of the shape of the cluster following properties and symmetry of the lattice under consideration. 
We would be able to approach the exact answer when we choose the clusters keeping the underlying symmetry in mind in a systematic way. 
The slow convergence for the case on the triangular lattice would be caused by an inadequate choice of the clusters on it.
This point should be clear in the future.

\section{Conclusion}
We predicted the accuracy thresholds of the topological color codes on the hexagonal and square-octagonal lattices, which are kinds of the quantum error correcting codes exploiting the topological properties in systems.
These accuracy thresholds correspond to the locations of the multicritical points for the random three-body Ising models on the triangular and Union-Jack lattices, respectively. 
The accuracy thresholds of these two topological color codes are expected to be lower than the probability for the quantum Gilbert-Varshamov bound with zero rate encoding. 
Accepting our estimations, we find that the accuracy thresholds of two color codes are equivalent to that of the quantum toric code on the square lattice, which corresponds to the location of the multicritical point for the $\pm J$ Ising model on the square lattice \cite{DKLP}. 
The accuracy threshold is given as $1-p_{c}=0.1092-3$ from several results by the improved method \cite{Ohzeki}, $1-p_{c}=0.10919(7)$ by highly precise Monte-Carlo simulation \cite{Hasen}, and a very recent estimation $1-p_{c}=0.10939(6)$ by the concept of the conformal field theory \cite{Queiroz}. 
In conclusion, the topological color code on the hexagonal and square-octagonal lattice would have equivalent robustness to that of the topological toric code on the square lattice despite their computational capabilities. 
This fact is indeed shown by the numerical simulation performed in very recent work by Katzgraber {\it et al.} \cite{CC}. 
The present work provided the same conclusion but would be valuable as the analytical predictions of the topological color codes on the hexagonal and square-octagonal lattices.

\begin{acknowledgments}
Fruitful discussions with H. G. Katzgraber and M. A. Martin-Delgado are greatly acknowledged. 
This work was partially supported by the Ministry of Education, Science, Sports and Culture, Grant-in-Aid for Young Scientists (B) under Grant No. 20740218, and for Scientific Research on the Priority Area \textquotedblleft Deepening and Expansion of Statistical Mechanical Informatics (DEX-SMI)\textquotedblright, and by CREST, JST.
\end{acknowledgments}


\end{document}